\documentclass[twocolumn,prb,aps,showpacs,superscriptaddress,amsfonts,amssymb,amsmath]{revtex4}
\usepackage{amsmath}
\usepackage{graphicx}
\begin{document}
\title{Absence of zero-temperature transmission rate of a double-chain tight-binding model for DNA
with random sequence of nucleotides in thermodynamic limit}
\author{Gang Xiong}
\affiliation{Physics Department, Beijing Normal University,
Beijing 100875, P. R. China}
\author{X. R. Wang}
\affiliation{Physics Department, The Hong Kong University of
Science and Technology, Clear Water Bay, Hong Kong SAR, P. R.
China}

\date{Draft on \today}

\begin{abstract}
The zero-temperature transmission rate spectrum of a double-chain
tight-binding model for real DNA is calculated. It is shown that a
band of extended-like states exists only for finite chain length
with strong inter-chain coupling. While the whole spectrum tends
to zero in thermodynamic limit, regardless of the strength of
inter-chain coupling. It is also shown that a more faithful model
for real DNA with periodic sugar-phosphate chains in backbone
structures can be mapped into the above simple double-chain
tight-binding model. Combined with above results, the transmission
rate of real DNA with long random sequence of nucleotides is
expected to be poor.
\end{abstract}

\maketitle

\begin{figure}[h]
 \includegraphics[height=2cm, width=4cm]{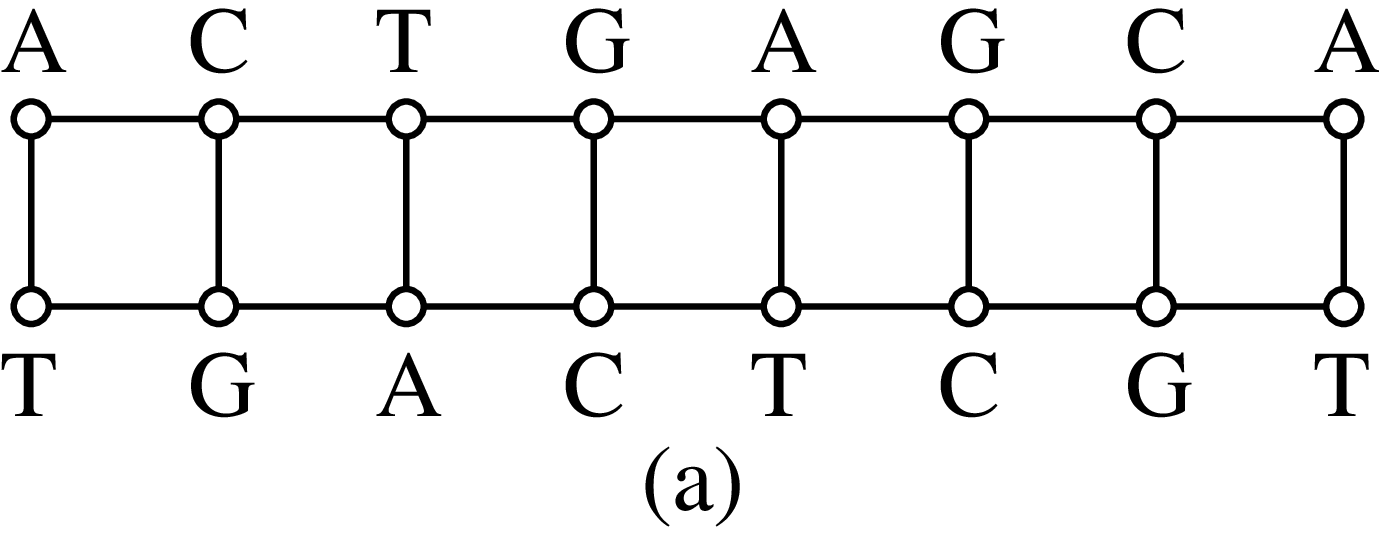}
 \includegraphics[height=2cm, width=4cm]{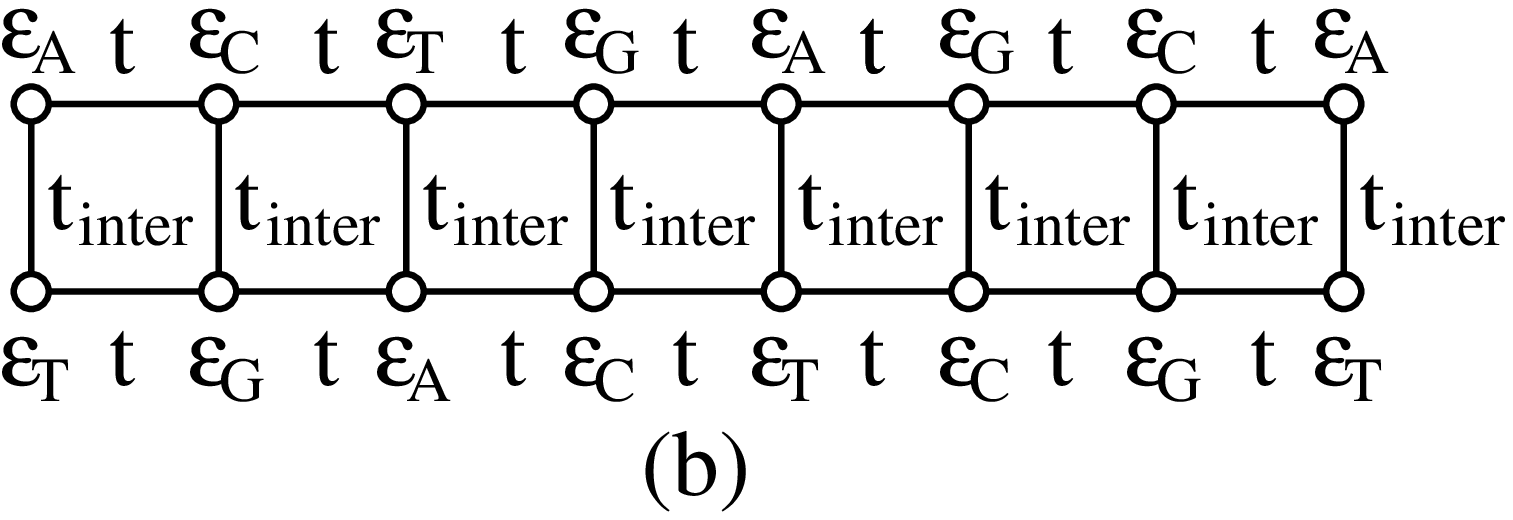}
\caption{(a)Schematic illustration of a part of DNA sequence;
(b)the corresponding double-chain TBM of the sequence in (a).
$\epsilon_A$, $\epsilon_T$, $\epsilon_C$, $\epsilon_G$ denote the
different on-site energies for the corresponding nucleotides. $t$
denotes intra-chain coupling and $t_{inter}$ inter-chain coupling.
}
 \label{dna}
\end{figure}

DNA is the biological molecule which keeps and propagates the
secretes of life for all the creatures on the earth, and it has
many interesting properties\cite{dekker}. It consists of two
chains of nucleotides. The nucleotides are of four types, usually
denoted as $A$, $T$, $C$ and $G$. Electrons are occupied on the
nucleotides and can hop from one nucleotide to its intra-chain and
inter-chain neighbors. Each pair of inter-chain nearest-neighbor
nucleotides can only belong to one of the four `base-pairs' $A-T$,
$T-A$, $C-G$ and $G-C$ while other kinds of combinations are
forbidden. Therefore, when the sequence of nucleotides in one
chain of a DNA is determined, the sequence in the other chain is
determined, too. Fig.\ref{dna}(a) shows schematically a part of a
DNA sequence.

The electronic transportation properties of DNA has attracted much
attention\cite{dekker} since Elley et. al. suggested that DNA may
become one-dimensional conductor\cite{elley}. Different
conclusions are obtained by different experiments. Some
experiments suggested that DNAs are conductors\cite{murphy}.
Later, other experiment groups claimed that DNAs are
insulators\cite{dunlap}. Direct measurement on single molecule
showed that they seem to be semiconductors\cite{porath}. Some
hints of superconduction behavior have even been
reported\cite{kasumov}. Further experiments show that the
conduction behavior of DNA molecules of different base-pair
sequences, say, identical base-pair sequence\cite{yoo} and
disordered base-pair sequence\cite{tran}, seems different.

In the aspect of theoretical studies, a simple one-electron model
was originally proposed by Iguchi\cite{iguchi1}. In this simple
model, each nucleotide is represented by a site and different
types of nucleotides have different on-site energies. Since the
coupling between nucleotides is short-range, it is assumed that
electron hopping occurs only between nearest-neighbor nucleotides
and the possible randomness in hopping constants is neglected.
Thus the electron behavior in a DNA is represented by a
double-chain tight-binding model(TBM). Many studies on electronic
properties of DNA are based on this model or its modified
versions\cite{song,yamada1,yamada2,yamada3,yamada4,roche,iguchi2}.
In this short paper, we shall show that this simple model may not
be appropriate for real DNAs if they do have conductor behavior. A
system should have extended states to provide conductor behavior.
It is well known by the famous Bloch theorem that a periodic TBM
has bands of extended states. However, the sequence of base-pairs
in each chain of real DNAs are {\it disordered} rather than {\it
periodic}, and this leads to random on-site energies in the
corresponding TBM. According to the scaling theory of
localization\cite{abrahams}, uncorrelated random on-site disorders
in a one-dimensional TBM will drive {\it all} electronic states to
exponentially localized states. At a first glance, the
double-chain TBM for DNAs seems different from a common TBM with
uncorrelated random on-site energies since the restriction of the
combination of inter-chain base-pairs in DNA will induce local
correlations between the two chains. However, by calculating the
zero-temperature transmission rate spectrum directly we shall show
that this double-chain TBM for real DNAs cannot have conductor
behavior, either. There have already been some theoretical studies
on similar
problems\cite{yamada1,yamada2,yamada3,yamada4,roche,iguchi2}, and
we shall also discuss and compare our results with them.

The Hamiltonian of the double-chain TBM for disordered DNAs is as
following
\begin{equation}
    \hat{H}_2=\hat{H}_{intra}+\hat{H}_{inter}
\end{equation}
where
\begin{eqnarray}
    \hat{H}_{intra}=\sum_n[\epsilon_{n,1}\hat{C}^{\dag}_{n,1}\hat{C}_{n,1}+\epsilon_{n,2}\hat{C}^{\dag}_{n,2}\hat{C}_{n,2}
    \nonumber\\
    +t(\hat{C}^{\dag}_{n+1,1}\hat{C}_{n,1}+\hat{C}^{\dag}_{n+1,2}\hat{C}_{n,2})]
\end{eqnarray}
and
\begin{equation}
    \hat{H}_{inter}=t_{inter}\sum_n\hat{C}^{\dag}_{n,1}\hat{C}_{n,2}.
\end{equation}
$\epsilon_{n,1}$ and $\epsilon_{n,2}$ are the on-site energies of
nucleotides in the two chains which can take four possible values
$\epsilon_A$, $\epsilon_T$,$\epsilon_C$,$\epsilon_G$. $t$ and
$t_{inter}$ are the intra-chain and inter-chain couplings,
respectively. Fig.\ref{dna}(b) shows the corresponding
double-chain TBM for the part of DNA sequence in Fig.\ref{dna}(a).

\begin{figure}[h]
 \includegraphics[height=2cm, width=4cm]{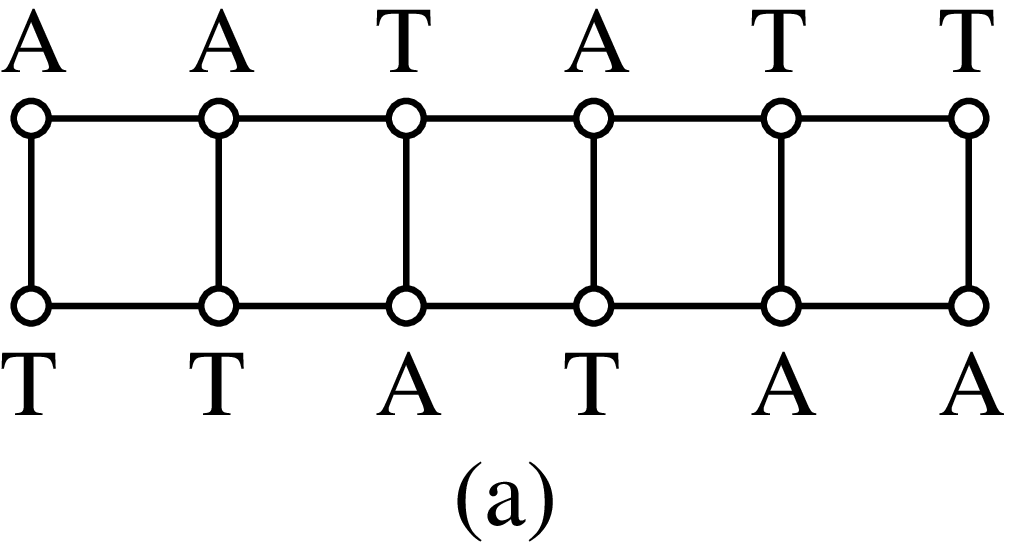}
 \includegraphics[height=2cm, width=4cm]{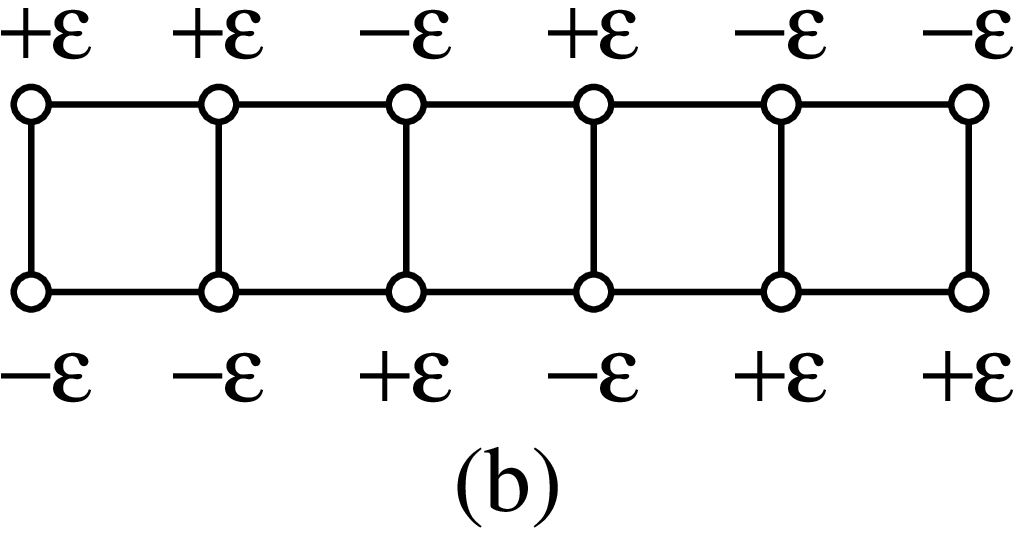}
\caption{The double-chain random tight-binding model for DNA.
(a)Schematic illustration of a part of the double-strand DNA with
only $A$ and $T$ as we consider; (b)the corresponding double-chain
tight-binding model.} \label{twochainmodel}
\end{figure}

In order to introduce the effect of disorder in on-site energies,
we consider the simplest case that only two types of nucleotides,
say, $A$ and $T$, exist in the two chains while the sequence of
$A$ and $T$ in each chain is random. This may be considered as a
minimum model to include on-site disorder effect because in a
general sequence in DNA all four possible base-pairs exist
randomly which will only enhance the effect of disorders. For this
special case, without loss of generality we can take the average
on-site energy of nucleotides $A$ and $T$ as the energy reference
point i.e., $(\epsilon_{A}+\epsilon_{T})/2=0$, and set the
intra-chain coupling $t$ as the energy unit, i.e., $t=1$. Then, we
need only consider the case that $\epsilon_A=+\epsilon$ and
$\epsilon_T=-\epsilon$, and $\epsilon$ can be considered as the
disorder strength in on-site energies. Fig.\ref{twochainmodel}(a)
shows a part of such a DNA sequence, and
Fig.\ref{twochainmodel}(b) is the corresponding TBM.

The zero-temperature transmission rate of an eigen-state of energy
$E$ in unit of $G_0=2e^2/h$ is given by \cite{imry}
\begin{equation}
    g(E)\equiv\frac{G(E)}{G_0}=\sum_{n}\frac{1}{1+\cosh\lambda_n(E)}
\end{equation}
where $\lambda_n(E)$ are eigenvalues of the operator
$\hat{T}^{\dag}(E)\hat{T}(E)$ and $\hat{T}(E)$ is the transfer
matrix of the system. The eigenvalues $\lambda_n(E)$ and the
transmission rate spectrum can be calculated by standard transfer
matrix algorithm\cite{tm}. This mathematical framework is
essentially the same as the one Roche et. al.\cite{roche} have
used in their study. We shall show that the transmission rate
spectrum of the double-chain TBM for disordered DNAs is strongly
suppressed and all electronic states are localized in the
thermodynamic limit.

\begin{figure}[h]
 \includegraphics[height=4cm, width=4cm]{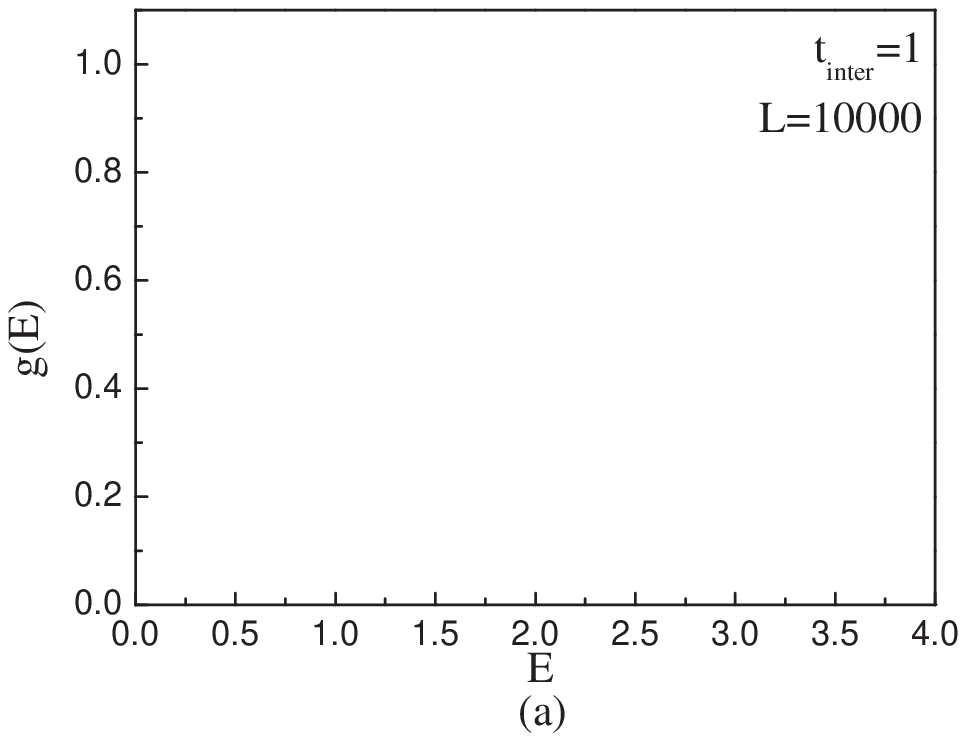}
 \includegraphics[height=4cm, width=4cm]{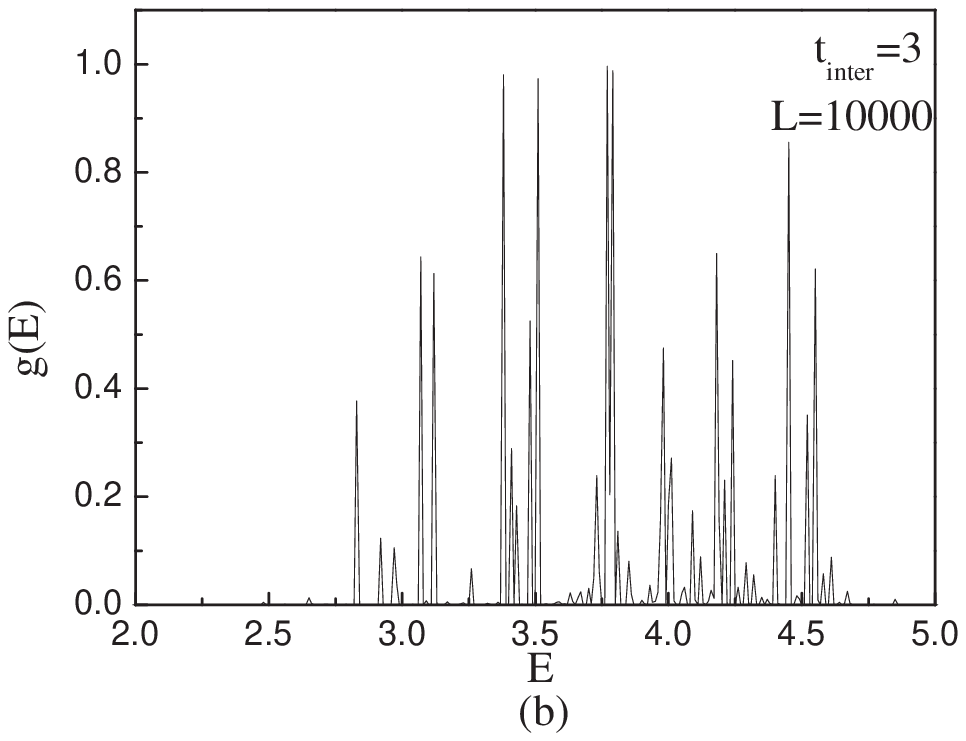}
 \includegraphics[height=4cm, width=4cm]{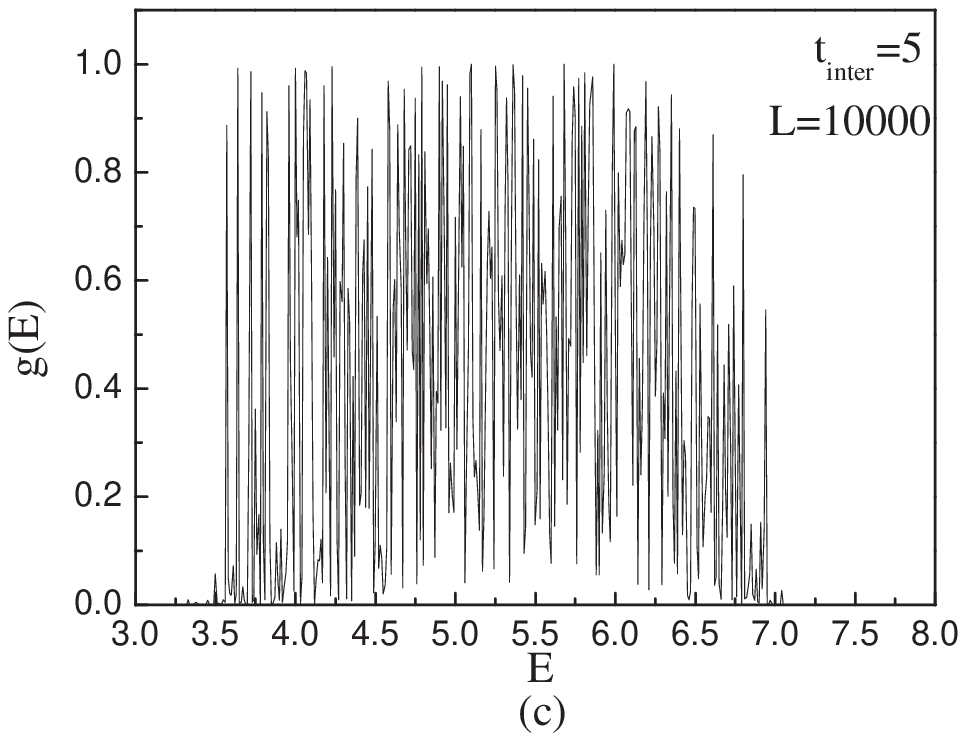}
 \includegraphics[height=4cm, width=4cm]{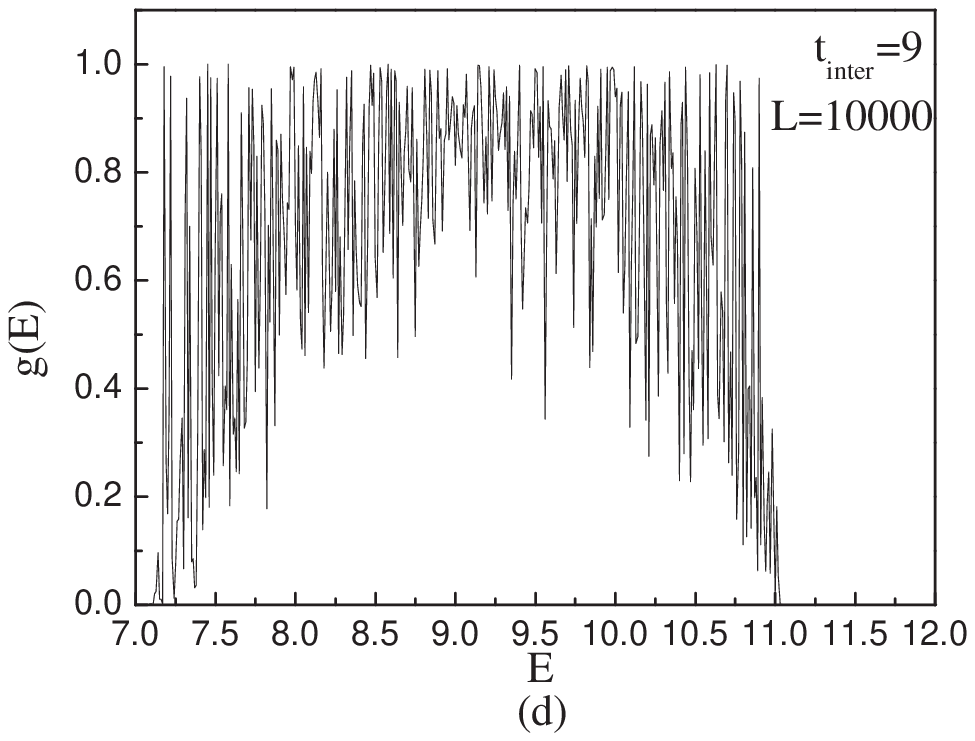}
 \includegraphics[height=4cm, width=4cm]{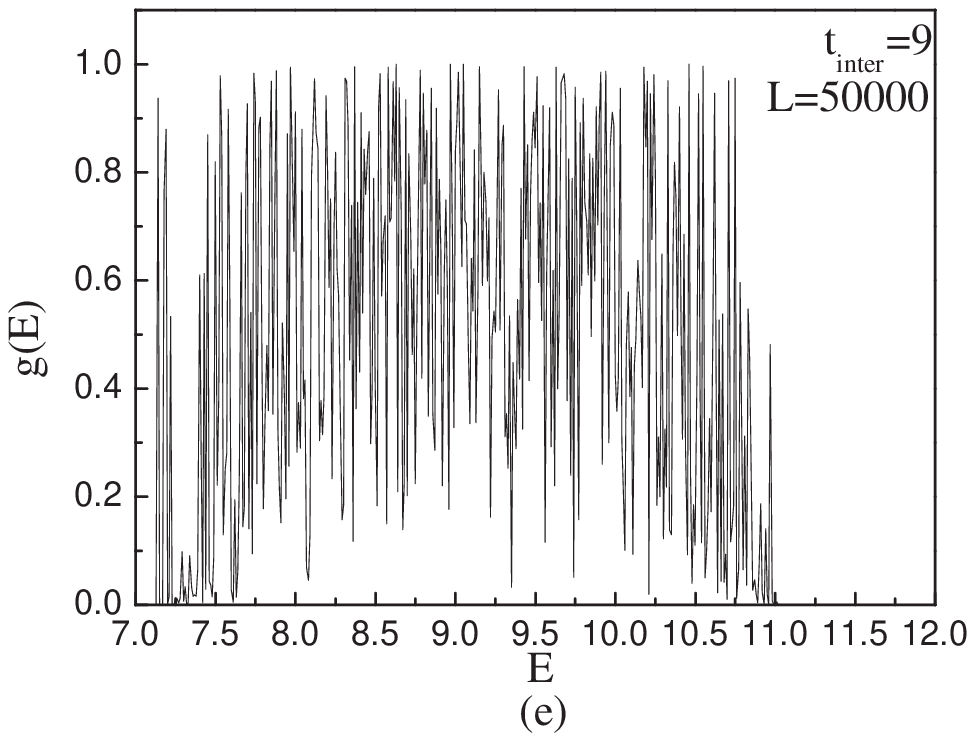}
 \includegraphics[height=4cm, width=4cm]{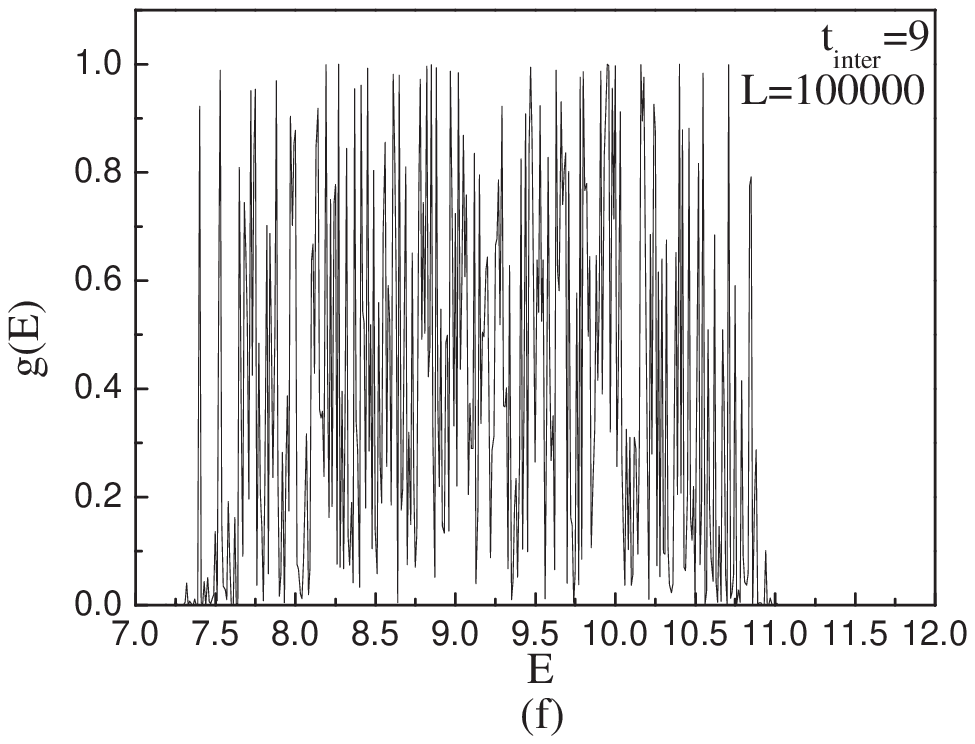}
\caption{The transmission rate spectrum g(E) for a double-chain
tight-binding model with random on-site energies. (a)$t_{inter}=1$
with $L=10000$, (b)$t_{inter}=3$ with $L=10000$,(c)$t_{inter}=5$
with $L=10000$, (d)$t_{inter}=9$ with $L=10000$, (e)$t_{inter}=9$
with $L=50000$, (f)$t_{inter}=9$ with $L=100000$.}
\label{twochain}
\end{figure}

Let us look at the numerical results. Since all results are
symmetric with E=0, we only provide the part of $E>0$.
Fig.\ref{twochain} is for $L=10000$ with (a)$t_{inter}=1$, (b)
$t_{inter}=3$, (c) $t_{inter}=5$, (d) $t_{inter}=9$. The parameter
for the disorder strength of on-site energies is taken as
$\epsilon=1$, and the probabilities of the occurrence of both $A$
and $T$ in each chain are set as $1/2$. Essentially the same
results are obtained for other non-zero value of $\epsilon$ and
non-zero occurrence probabilities of $A$ and $T$.

For weak inter-chain coupling $t_{inter}=1$, the transmission rate
is zero everywhere, which means that all states are localized.
When inter-chain coupling increases, say, $t_{inter}=3$, peaks
begin to exist in the spectrum, similar to the behavior of a
random single-chain TBM of finite length. For further strong
inter-chain coupling, say, $t_{inter}=5$, the curve begins to form
a band with sharp peaks. With further increase of $t_{inter}$,
say, $t_{inter}=9$, a band of finite transmission rate emerges.
However, as shown in Fig.\ref{twochain}(d), (e) and (f) for
$t_{inter}=9$ with $L=10000$, $50000$ and $100000$, respectively,
it is clear that this band tends to vanish at large $L$. Thus the
existence of a band of finite transmission rate in the case of
strong inter-chain coupling does not mean the existence of truly
extended states. It is only a finite-size effect which is expected
to disappear when the chain length $L$ tends to infinity.

According to the above results, we may come to the following
conclusions. A disordered double-chain TBM for real DNAs with weak
inter-chain couplings does not have conducting electronic states.
When inter-chain couplings become strong, the model with finite
chain length forms an energy band of finite transmission rate,
similar to that of a periodic single-chain TBM. However, this is
due to finite-size effect and does not mean the existence of true
conducting states. In the thermodynamic limit, the band tends to
disappear and all electronic states are localized. Therefore, the
conductor behavior of real DNAs, if exists, cannot be explained by
the simple one-electron double-chain TBM with random on-site
energies.

\begin{figure}[h]
 \includegraphics[height=3cm, width=4cm]{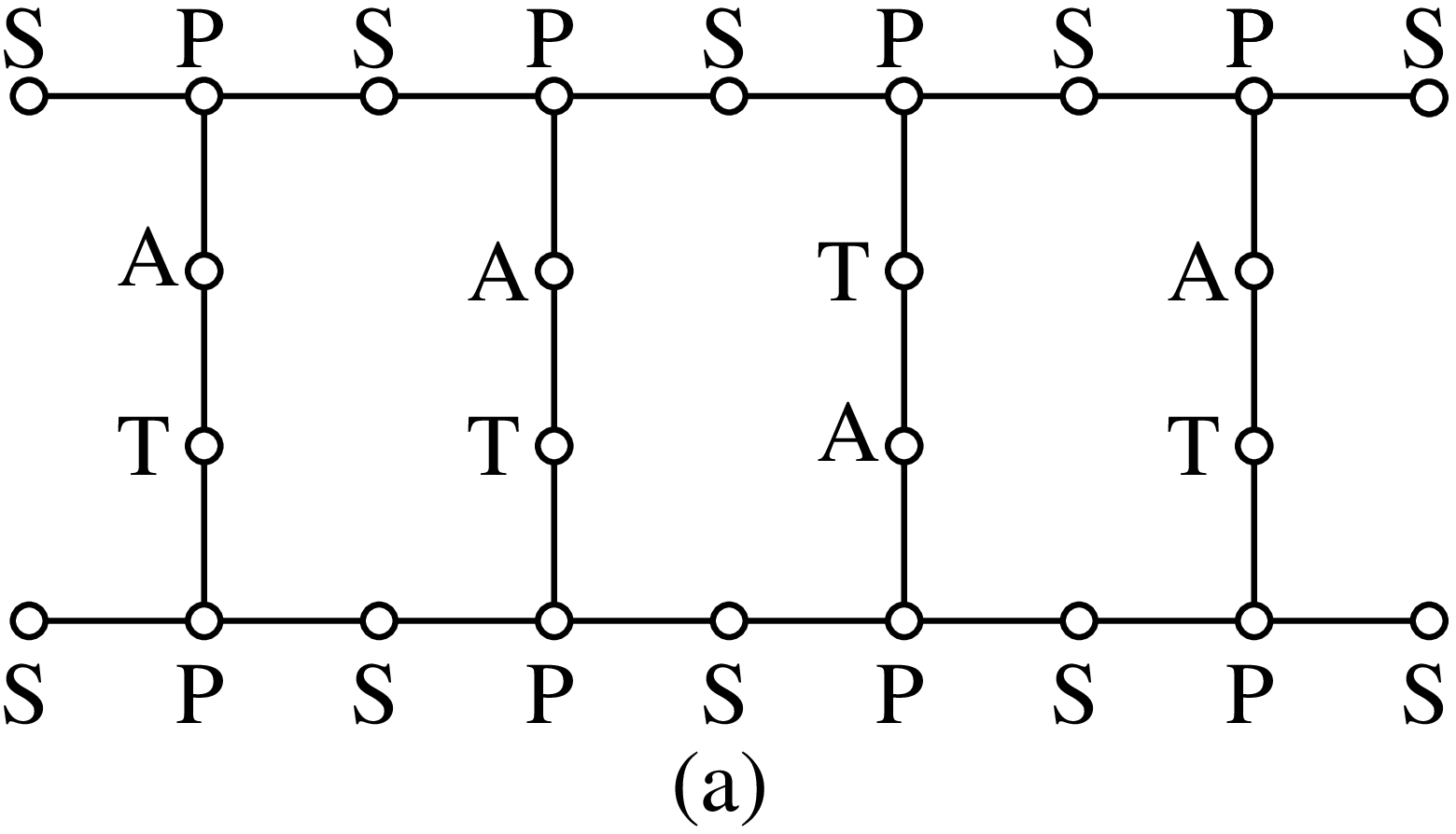}
 \includegraphics[height=3cm, width=4cm]{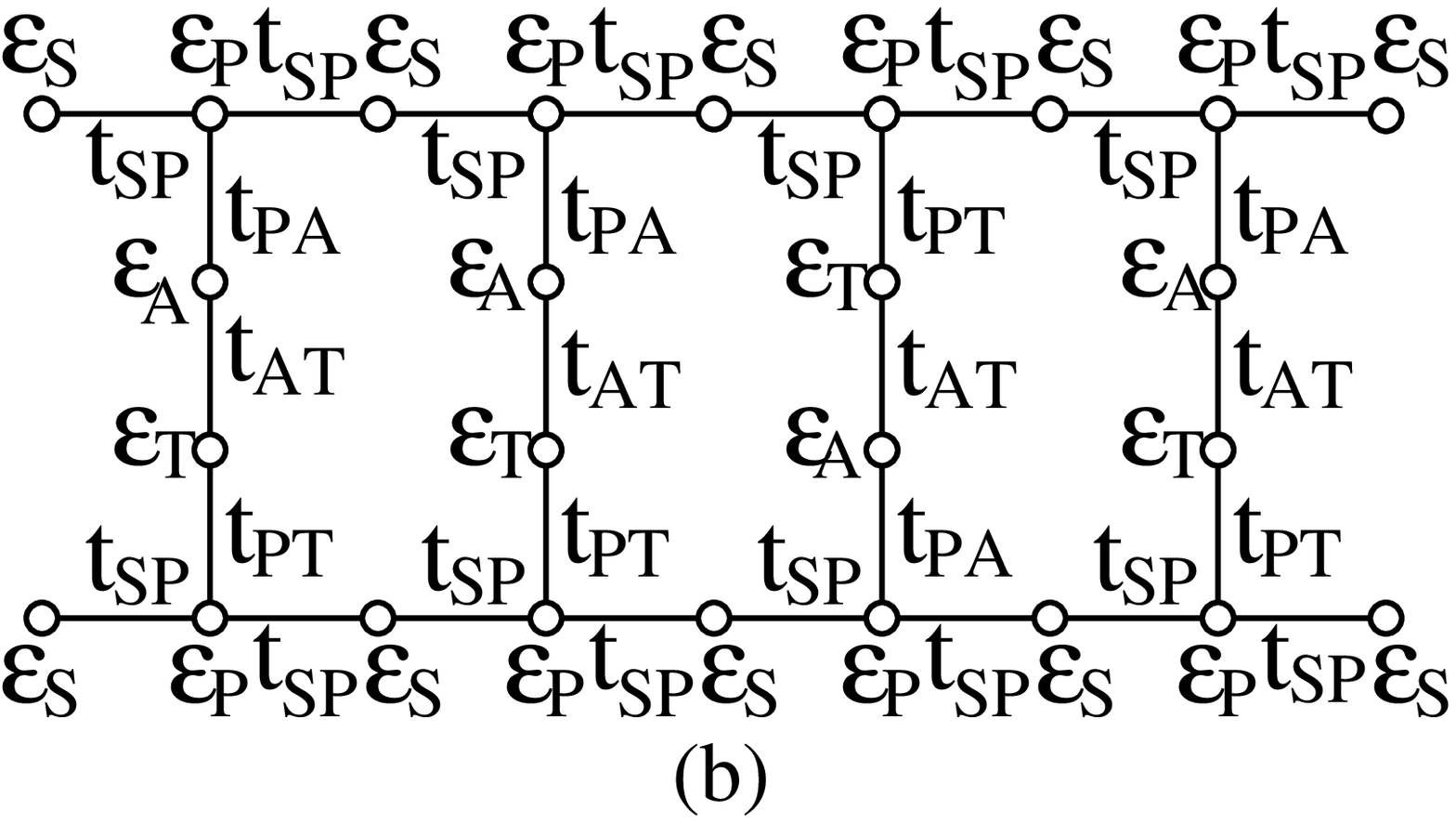}
\caption{(a)Schematic illustration of the faithful model of a part
of double-chain DNA with $A$ and $T$ nucleotides only; (b)the
corresponding double-chain TBM of the sequence in (a). $S$ and $P$
denote sugar and phosphate sites. $\epsilon_S$ and $\epsilon_P$
denote on-site energies of sugar and phosphate sites,
respectively. $\epsilon_A$ and $\epsilon_T$ denote on-site
energies of $A$ and $T$ nucleotides, respectively. $t_{SP}$
denotes intra-chain coupling between a sugar and its neighboring
phosphate. $t_{PA}$ and $t_{PT}$ the coupling between a phosphate
and an $A$ and $T$, respectively.} \label{twochainmodel02}
\end{figure}
However, it is well-known that there is a periodic sugar-phosphate
chain in each backbone structure of real DNA. Therefore, a
faithful model for a real double-chain DNA should be as shown in
Fig.\ref{twochainmodel02}(a) and (b), where $S$ and $P$ denote
sugar and phosphate and $A,T$ denote the nucleotides. Thus there
comes a question on whether the simple model in Fig.\ref{dna}
without sugar-phosphate periodic chains can be used for real DNA.
Fortunately, we shall show below that the model in
Fig.\ref{twochainmodel02} can be renormalized into the model in
Fig.\ref{twochainmodel} by the real-space decimation
renormalization technique\cite{wang}. We shall still restrict to
the simplest case that only $A$ and $T$ exist in the two chains.
\begin{figure}[h]
 \includegraphics[height=2cm, width=4cm]{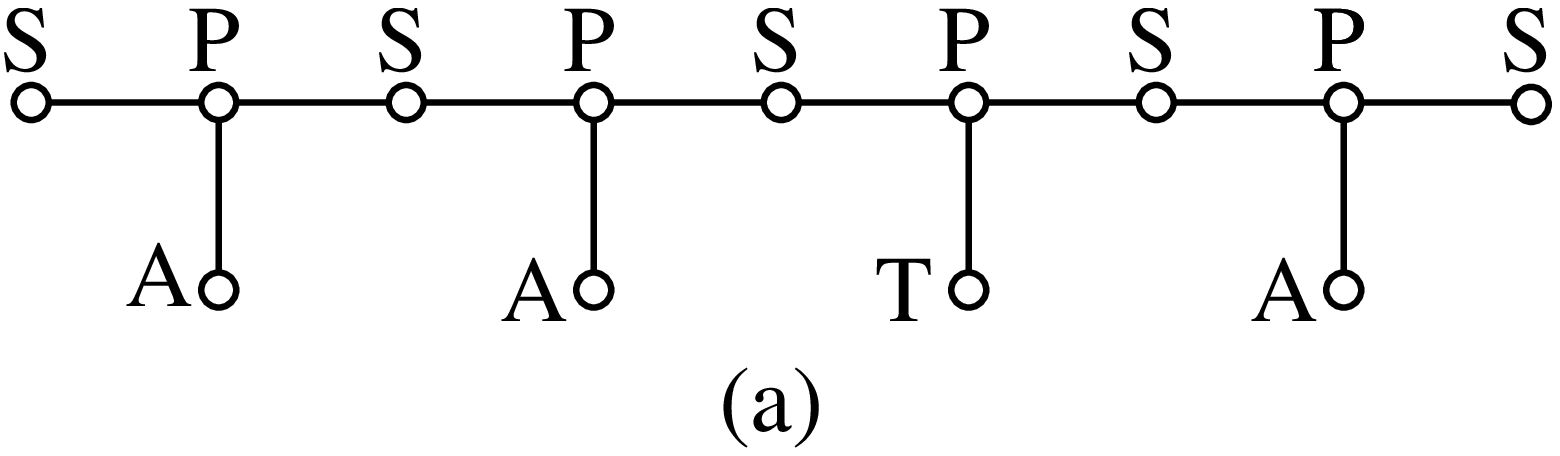}
 \includegraphics[height=2cm, width=4cm]{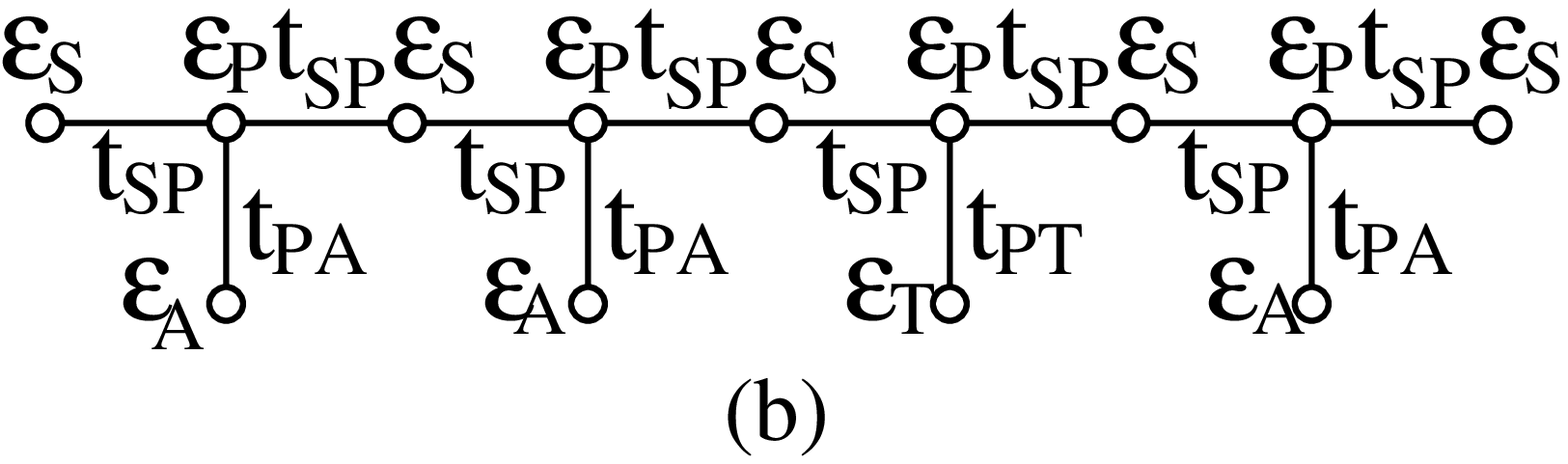}
\caption{(a)Schematic illustration of the faithful model of a part
of a single-chain DNA with only $A$ and $T$ nucleotides; (b)the
corresponding single-chain TBM of the sequence in (a). $S$ and $P$
denote sugar and phosphate sites. $\epsilon_S$ and $\epsilon_P$
denote on-site energies of sugar and phosphate sites,
respectively. $\epsilon_A$ and $\epsilon_T$ denote on-site
energies for $A$ and $T$ nucleotides, respectively. $t_{SP}$
denotes intra-chain coupling between a sugar and its neighboring
phosphate. $t_{PA}$ and $t_{PT}$ the coupling between a phosphate
and $A$ and $T$ nucleotides, respectively.}
\label{onechainmodel02}
\end{figure}

\begin{figure}[h]
 \includegraphics[height=2cm, width=5cm]{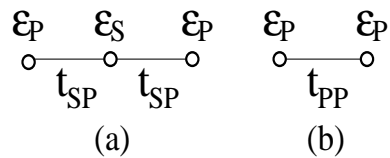}
\caption{A local phosphate-sugar-phosphate structure (a) is
renormalized into a phosphate-phosphate structure (b) by
decimation of the sugar site S.} \label{renormalize01}
\end{figure}

Let us first look at a single-chain DNA with series of random
nucleotides as shown in Fig.\ref{onechainmodel02}(a). We shall
show that it can be mapped into a single chain Anderson TBM with
random on-site energies. As a first step, let us renormalize a
local phosphate-sugar-phosphate structure into a
phosphate-phosphate structure by decimation of the sugar site, as
shown in Fig.\ref{renormalize01}. Perform this decimation process
on all local phosphate-sugar-phosphate structures, one can
decimate all sugar sites. The decimation process renormalizes the
on-site energy of all phosphate sites into $\epsilon_P^{\prime}$
\begin{eqnarray}
    \epsilon_P^{\prime}=\epsilon_P+\frac{2t_{SP}^{2}}{E-\epsilon_{S}}
\label{onsite}
\end{eqnarray}
and introduces a direct coupling $t_{PP}$ between
nearest-neighboring phosphates
\begin{eqnarray}
    t_{PP}=\frac{t_{SP}^{2}}{E-\epsilon_{S}}.
\end{eqnarray}
$E$ is the eigen-energy we consider, $\epsilon_S$ and $\epsilon_P$
are on-site energies of sugar and phosphate sites, and $t_{SP}$ is
the coupling between neighboring sugar and phosphate sites in the
original model of Fig.\ref{onechainmodel02}. (Both
$\epsilon_P^{\prime}$ and $t_{PP}$ diverge at $E=\epsilon_S$ which
should be treated by considering directly the original stationary
Schrodinger equations.)  For each given $E$, the above decimation
only gives a global shift for the on-site energies of each
phosphate site and does not introduce any disorder because
$\epsilon_S$, $\epsilon_P$ and $t_{SP}$ are constant values in the
original model.

\begin{figure}[h]
 \includegraphics[height=2cm, width=4cm]{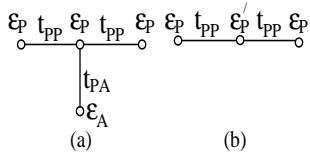}
\caption{A local phosphate-phosphate-phosphate structure with the
middle phosphate coupled with a nucleotide (a) is renomalized into
a phosphate-phosphate-phosphate structure structure (b) by
decimation of the nucleotide coupled with the middle P.}
\label{renormalize02}
\end{figure}

The second step is to renormalize a local
phosphate-phosphate-phosphate structure by decimation of the
nucleotide coupled with the middle phosphate as shown in
Fig.\ref{renormalize02}. This decimation process only renormalizes
the on-site energy of the middle phosphate $\epsilon_P$ into
\begin{eqnarray}
    \epsilon_P^{\prime}=\epsilon_P+\frac{t_{PA}^2}{E-\epsilon_A}
\end{eqnarray}
where $\epsilon_A$ is the on-site energy of the nucleotide coupled
with the middle phosphate. ($\epsilon_P^{\prime}$ diverges at
$E=\epsilon_A$, which means that the electron wave of energy
$E=\epsilon_A$ is totally reflected at this point, similar to the
`anti-resonance' phenomenon\cite{antiresonance}). Since the series
of $A$ and $T$ nucleotide are random for the case we consider,
this decimation process introduces effectively randomness into
renormalized on-site energies of phosphate sites. Perform this
decimation step for all local phosphate-phosphate-phosphate
structures, one finally obtain a single chain of phosphate sites
with uncorrelated random on-site energies. According to the
scaling theory\cite{abrahams}, the on-site disorder localizes
states of all eigen-energies.

\begin{figure}[h]
 \includegraphics[height=3cm, width=5cm]{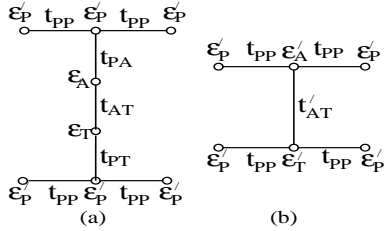}
\caption{A local structure (a) is renormalized into a local
structure (b) by decimation of the nucleotide pair.}
\label{renormalize03}
\end{figure}

Now let us consider the double-chain model as shown in
Fig.\ref{twochainmodel02}(b). Perform the same process as shown in
Fig.\ref{renormalize01}, one can decimate all sugar sites. In
order to map into the model of Fig.\ref{twochainmodel}, one needs
to decimate all the nucleotide pairs. This can be done by
considering the decimation process shown in
Fig.\ref{renormalize03} where a pair of coupled nucleotides
connecting two phosphates in each chain are decimated. This will
renormalize the on-site energies of the two phosphates as
\begin{eqnarray}
    \epsilon_A^{\prime}=\epsilon_P^{\prime}+\frac{(E-\epsilon_T)t_{PA}^2}
    {(E-\epsilon_A)(E-\epsilon_T)-t_{AT}^2}\nonumber\\
    \epsilon_T^{\prime}=\epsilon_P^{\prime}+\frac{(E-\epsilon_A)t_{PT}^2}
    {(E-\epsilon_A)(E-\epsilon_T)-t_{AT}^2}
\end{eqnarray}
($\epsilon_P^{\prime}$ is as in Eq.(\ref{onsite})) and introduce a
direct inter-chain coupling between the two phosphates as
\begin{eqnarray}
    t_{AT}^{\prime}=\frac{t_{PA}t_{PT}t_{AT}}
    {(E-\epsilon_A)(E-\epsilon_T)-t_{AT}^2}.
\end{eqnarray}
($\epsilon_A^{\prime}$, $\epsilon_T^{\prime}$ and
$t_{AT}^{\prime}$ diverge at
$(E-\epsilon_A)(E-\epsilon_T)-t_{AT}^2=0$ which should be treated
by considering directly the original stationary Schrodinger
equations.) It is easy to see that this decimation process
introduces correlated randomness into on-site energies of each
pair of inter-chain-coupled phosphate sites. Therefore, the model
in Fig.\ref{twochainmodel02} is mapped into essentially the same
model as shown in Fig.\ref{twochainmodel}. According to numerical
results for the model in Fig.\ref{twochainmodel}, we may conclude
that no extended states exist in thermodynamic limit for both
models.
\begin{figure}[h]
 \includegraphics[height=3cm, width=4cm]{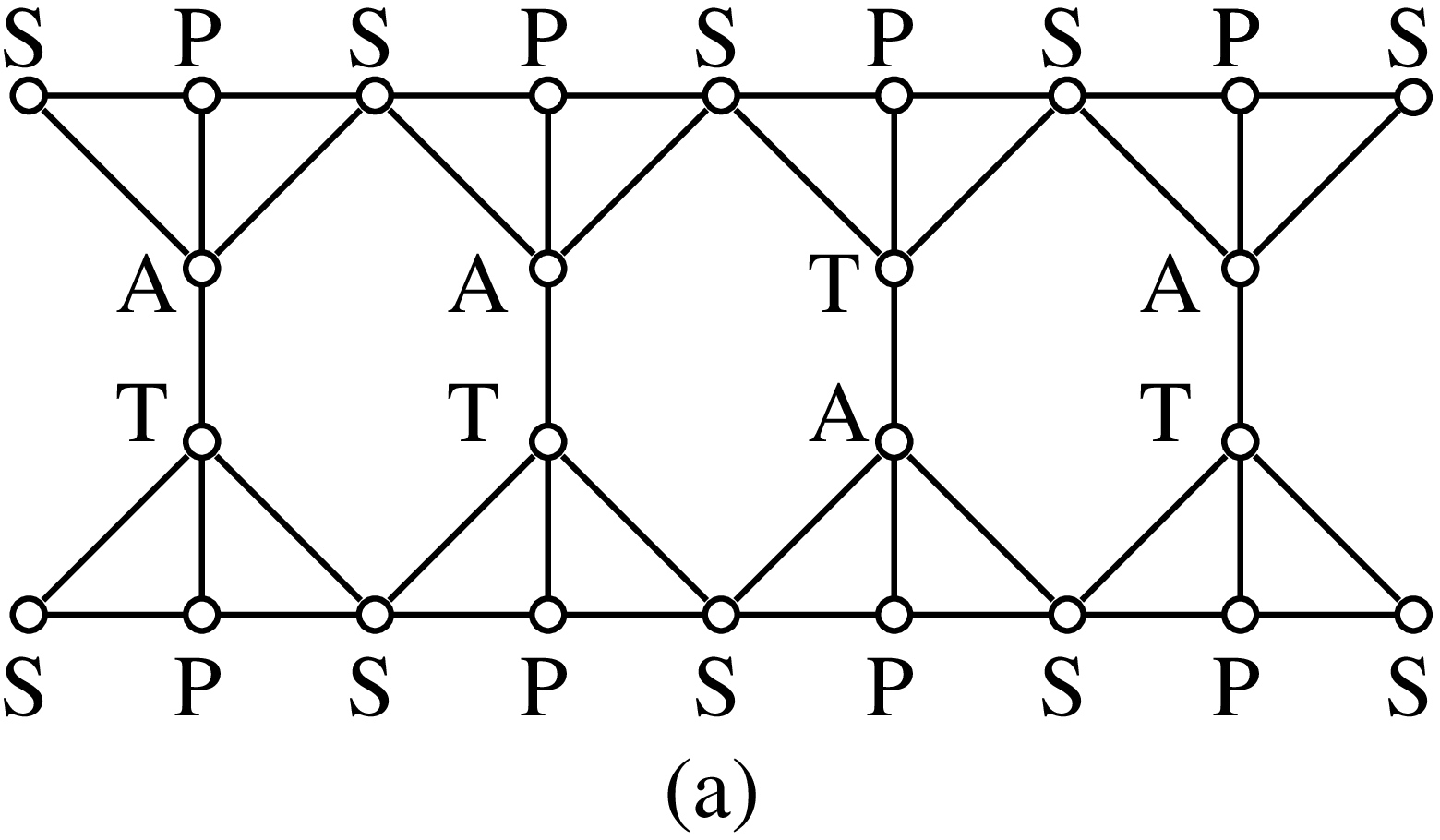}
 \includegraphics[height=2cm, width=4cm]{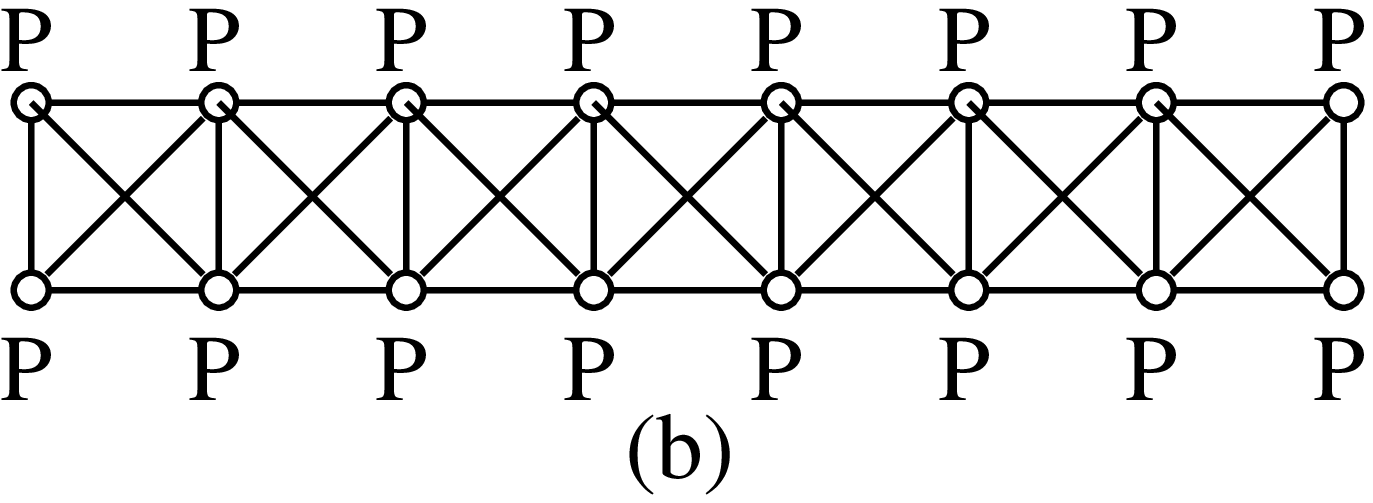}
\caption{A double-chain TBM model for DNA with couplings between
nucleotide and its neighboring sugar sites (a) is renormalized
into a double-chain TBM (b) by decimation of the sugar sites and
nucleotide pairs.} \label{twochainmodel03}
\end{figure}

It should be noted that extended states may emerge when other
kinds of couplings are included in the above model. For example,
let us take into account the couplings between a nucleotide and
its nearest-neighboring sugar sites, as shown in
Fig.\ref{twochainmodel03}(a). Then, by decimation of both the
sugar sites and nucleotide pairs, one can obtain a model of
Fig.\ref{twochainmodel03}(b) where both random on-site energies
and random couplings in each group of four nearest-neighboring
sites are correlated. Such a model may have extended states
because it looks similar to generalized dimer models which have
been shown to have extended states in recent studies\cite{dimer}.

Before making the summary, we would like to make some discussions
about our results and recent
theoretical\cite{roche,iguchi2,yamada1,yamada2,yamada3,yamada4}
and experimental \cite{tran,yoo}results on similar problems.
Iguchi\cite{iguchi2} has studied a single chain of DNA with
periodic nucleotide series, which always has extended states due
to Bloch's theorem. Roch et. al.\cite{roche} have studied both
periodic approximations of aperiodic DNA series and series
extracted from real DNAs. Their results for series of real DNAs
are similar with our results. Yamada\cite{yamada1,yamada2,yamada3}
and Yamada et. al.\cite{yamada4} have studied the influence of
correlated on-site disorder in the model and found similar
localization behaviors as our results suggest. In the aspect of
experiments, Yoo et. al.\cite{yoo} have found that the conduction
behavior of DNA with identical base-pairs may be well explained by
bands of extended states separated by localized states which is
possible for a periodic TBM model. While Tran et. al.'s
studies\cite{tran} on $\lambda-$DNA with disordered sequence of
base-pairs seem to suggest that no truly extended states exist,
which agrees with our results.

In summary, we have studied the disordered double-chain TBM with
and without periodic sugar-phosphate chain in the backbone
structure for real DNAs. Numerical results combined with
real-space decimation renormalization technique show that in both
cases no truly extended states exist in the thermodynamic limit,
regardless of how strong the inter-chain couplings are. Therefore,
the double-chain TBM for real DNAs considered in this paper always
has zero transmission rate in thermodynamic limit, and bad
conduction is expected for real DNAs with long random series of
nucleotides.

GX acknowledges the support of CNSF under grant No. 10347101 and
the Grant from Beijing Normal University.

\end{document}